\documentclass[singlecolumn,amssymb, nobibnotes, aps, preprint,nopacs,superscriptaddress, groupedaddress]{revtex4}

\usepackage{amsmath}
\usepackage[dvips]{graphicx}
\usepackage{epsfig}
\usepackage{tabularx}
\usepackage{color}
\usepackage[pdfpagemode=UseNone,colorlinks=true,linkcolor=blue,citecolor=blue]{hyperref}

\usepackage{soul}
\usepackage{gensymb}

\newcommand{\OmegaX}{\Omega_{\scriptstyle{x}}}
\newcommand{\OmegaY}{\Omega_{\scriptstyle{y}}}
\newcommand{\OmegaZ}{\Omega_{\scriptstyle{z}}}

\newcommand{\gX}{g_{\scriptscriptstyle{x}}}
\newcommand{\gY}{g_{\scriptscriptstyle{y}}}

\newcommand{\GX}{\Gamma_{\scriptscriptstyle{x}}}
\newcommand{\GY}{\Gamma_{\scriptscriptstyle{y}}}

\newcommand{\matV}{\mathbf{V}}

\newcommand{\ud}{\mathrm{d}}
\newcommand{\chic}{\chi_{\scriptscriptstyle{c}}}
\newcommand{\Qc}{Q_{\scriptscriptstyle{c}}}
\newcommand{\Pc}{P_{\scriptscriptstyle{c}}}

\begin{document}

\title[]{Optical squeezing mediated by levitated oscillators at their quantum ground state}

\author{A. Pontin}
\email{antonio.pontin@cnr.it}
\affiliation{CNR-INO, largo Enrico Fermi 6, I-50125 Firenze, Italy}

\author{Q. Deplano}%
\affiliation{Dipartimento di Fisica e Astronomia, Univ. Di Firenze, via Sansone 1, I-50019 Sesto Fiorentino, Italy}%
\affiliation{INFN, Sezione di Firenze, via Sansone 1, I-50019 Sesto Fiorentino, Italy}

\author{F. Marino}%
\affiliation{CNR-INO, largo Enrico Fermi 6, I-50125 Firenze, Italy}
\affiliation{INFN, Sezione di Firenze, via Sansone 1, I-50019 Sesto Fiorentino, Italy}

\author{F. Marin}
\email{francesco.marin@unifi.it}
\affiliation{CNR-INO, largo Enrico Fermi 6, I-50125 Firenze, Italy}
\affiliation{Dipartimento di Fisica e Astronomia, Univ. Di Firenze, via Sansone 1, I-50019 Sesto Fiorentino, Italy}%
\affiliation{INFN, Sezione di Firenze, via Sansone 1, I-50019 Sesto Fiorentino, Italy}
\affiliation{European Laboratory for Non-Linear Spectroscopy (LENS), Via Carrara 1, I-50019 Sesto Fiorentino, Italy}

\begin{abstract}
We demonstrate optical squeezing below the shot-noise level generated through the interaction of an optical cavity field with two center-of-mass modes of a levitated nanoparticle, simultaneously cooled to occupation numbers well below unity. By analyzing the quadrature fluctuations of the cavity output through heterodyne detection, we resolve the full spectral covariance matrix of the optical field and map regions of sub-shot-noise squeezing as a function of detection phase and frequency. Operating in the resolved sideband and strong coupling regime where mechanical modes hybridize with the optical mode, we observe consistent squeezing in the band 70-95 kHz with a lowest variance of 0.98 (2$\%$ below vacuum fluctuations). We thus demonstrate optical squeezing mediated by multiple mechanical oscillators in their quantum ground state, bridging mechanical quantum control with non-classical light and establishing levitated optomechanics as a platform for multimode quantum interactions.
 
\end{abstract}

\maketitle
At the dawn of cavity optomechanics the two major experimental milestones, which remained elusive for a long time, were the ground state cooling of a macroscopic mechanical oscillator and the generation of so-called ponderomotive squeezing. The latter arises when quantum correlations in the quadratures of an optical field, induced by optomechanical interaction, allow the achievement of sub shot noise fluctuations~\cite{Fabre1994Quantumnoise,Mancini1994Quantumnoise}. These two objectives have since been accomplished~\cite{Chan2011laser,purdy2013strong}, but the fact remains that they have mutually exclusive favorable regimes:  motional cooling is enhanced in the resolved sidebands regime~\cite{Marquardt2007Quantum}, while ponderomotive squeezing is better observed in the bad-cavity regime~\cite{Fabre1994Quantumnoise}.  The field has since significantly progressed, with the observation of a range of quantum phenomena, from the measurement of an oscillator motion near the Heisenberg limit~\cite{Rossi2018Measurementbased}, to multimode~\cite{Nielsen2016Multimode} and room temperature~\cite{Huang2024Roomtemperature} squeezing, and even more deeply non-classical phenomena~\cite{Chen2020Entanglement,Bild2023Schrodingercat}.

Despite these advances, a key challenge remains largely unaddressed: the simultaneous observation of quantum behavior in both the mechanical and optical subsystems. Most demonstrations of ponderomotive squeezing have not prepared a quantum mechanical oscillator, as indeed a classical oscillator suffices. The observation of squeezing mediated by a mechanical mode that is cooled well below unitary occupation represents a genuine advance toward understanding the rich interplay between quantum optical fields and mechanical quantum systems.

Here, we demonstrate optical squeezing below shot noise, mediated by the interaction between an optical cavity field and two center-of-mass modes of a levitated nanoparticle, each cooled to occupation numbers substantially below unity. 
Levitated optomechanics has recently emerged as a versatile platform and a promising avenue for sensing, both for fundamental science~\cite{Monteiro2020Search,Kilian2024Darkmatter,Aggarwal2022Searching} and technological applications~\cite{Rademacher2019Quantum}. In the context of the two optomechanical milestones discussed above, ground state cooling has been achieved for a single~\cite{Delic2020science,Magrini2021,novotny2021_GS,Ranfagni2021Two-dimensional,Kamba2022Optical,Dania2025Highpurity} and two oscillating modes~\cite{Piotrowski2023Simultaneous,Deplano2025highpurity} and squeezed light has been observed~\cite{Militaru2022Ponderomotive,Magrini2022Squeezed}, notably, without using an optical cavity.
The demonstration of simultaneous quantum behavior in both mechanical and optical subsystems, reported in this work, highlights the platform readiness for exploitation as a true quantum resource.  It represents a significant step towards the generation of entanglement~\cite{bose1997Preparation,vitali2007Optomechanical,paternostro2007Creating,Deplano2026Stationary}  and the use of mechanical oscillators as local quantum memories~\cite{Mancini2003Scheme,Pirandola2006Macroscopic}.

\begin{figure}[h]
    \centering
    \includegraphics[width=8.6cm]{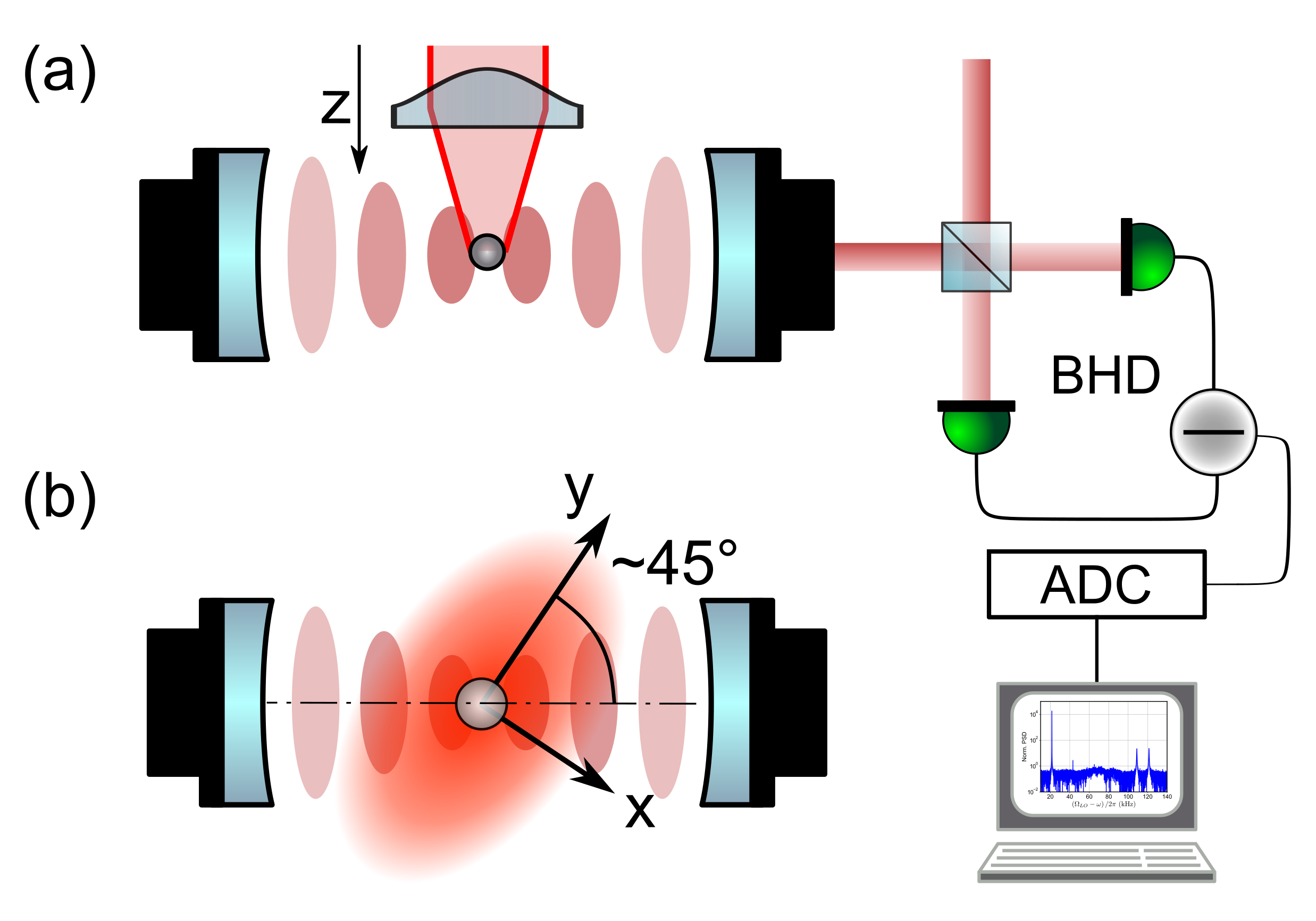}
    \caption{Pictorial view of the experiment. a) An optical tweezer confines a silica nanosphere. The particle is positioned at the center of an optical cavity, near a node of the intracavity standing wave.
    The cavity output field is analyzed by a balanced heterodyne detection (BHD) whose output is acquired through an analog-to-digital converter (ADC). b) The tweezer focus is slightly elliptical in the transverse plane with major ($y$) and minor ($x$) axes rotated by $\sim45\degree$ from the cavity axis.}
    \label{Fig0}
\end{figure}

The experimental system on which this work is based is presented in Ref.~\cite{Deplano2025highpurity}, where the focus is on the properties of the mechanical subsystem. Here we analyze the same experimental data to derive the quantum properties of the optical field at the output of the cavity. Crucially, the system operates in the resolved sideband regime, which is much less favorable for the observation of squeezing, and in the strong coupling regime where optical and mechanical modes hybridize.  By analyzing the quadrature fluctuations of the cavity output field through heterodyne detection, we resolve the full spectral covariance matrix of the optical field and map regions of sub-shot-noise squeezing as a function of detection phase and frequency. These results expand the understanding of quantum optomechanical systems beyond single-mode dynamics and demonstrate a new regime where collective quantum behavior of multiple mechanical modes shapes the optical field properties.

\emph{Experimental setup}  - A schematic view of the setup is shown in Fig.~\ref{Fig0}. It is based on a coherent scattering approach~\cite{Vuletic2000Laser,Delic2020science} where a silica nanoparticle, confined in an optical tweezer, is placed at the center of an empty high-finesse cavity. Light scattered by the particle populates the cavity mode when the tweezer field is resonant with the cavity. The resulting interference creates a large coupling between the particle motion and the cavity field~\cite{toros2020Quantum,toros2021Coherent}. A detailed description of the experimental setup can be found in Refs.~\cite{Ranfagni2021Vectorial, Deplano2025highpurity}, while here we summarize only its main aspects.  

The optical tweezer provides a 3D trapping potential with frequencies $(\OmegaX,\OmegaY,\OmegaZ)/2\pi\simeq(121,109,21)$\,kHz. The motion in the transverse plane of the tweezer, $x-y$, is coupled to the cavity field with rates $\gX$ and $\gY$, respectively, and is heated mainly by residual gas collisions and photon recoil. Their total effect is captured by the heating rates $\GX$ and $\GY$.  The motion along the tweezer propagation axis, $z$, can be safely neglected, since it is very weakly coupled to the cavity mode and there is a sufficiently large separation between mechanical resonances. The optical cavity has a full linewidth $\kappa/2\pi=57$\,kHz, as such the transverse dynamics is well within the resolved sidebands regime ($\kappa\ll\Omega_i$)~\cite{Marquardt2007Quantum}. Thus, setting an optimal detuning $-\Delta\simeq\Omega_i$ the cavity provides efficient motional cooling~\cite{cavity2014aspelmeyer}.  The cavity output field is analyzed through a balance heterodyne detection (BHD). The parameters that are experimentally determined, as reported in Ref.~\cite{Deplano2025highpurity}, are summarized in Tab.~\ref{tab:par}. These allowed to estimate mean occupation numbers of $n_x=0.55\pm0.03$ and $n_y=0.74\pm0.04$, both well below the unity threshold.

\begin{table}[h]
    \centering
    \begin{tabular}{lccc}
        \hline
        $\gX / 2\pi$ & $\gY / 2\pi$ &
        $\GX / 2\pi$ & $\GY / 2\pi$ \\
        \hline
        $14.13 \pm 0.22$ & $10.37 \pm 0.16$ &
        $4.03 \pm 0.32$  & $3.05 \pm 0.26$ \\
        \hline
    \end{tabular}
    \caption{System parameters determined in Ref.~\cite{Deplano2025highpurity}: coupling rates $g_i$ between the cavity mode and particle motion in the tweezer transverse plane $x-y$, and the associated heating rates $\Gamma_i$. All quantities are expressed in kHz.}
    \label{tab:par}
\end{table}

\begin{figure*}[h]
    \centering
    \includegraphics[width=\textwidth]{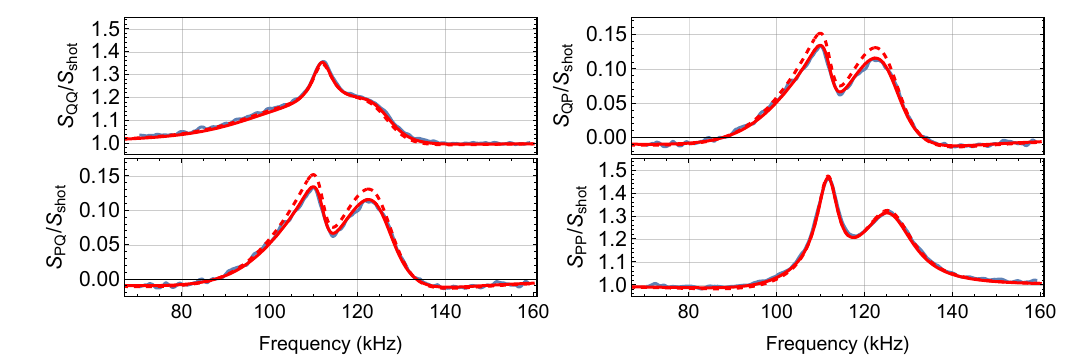}
    \caption{Spectral covariance matrix. Fluctuations of the cavity output field quadratures are reconstructed from a balanced heterodyne measurement through a phase sensitive analysis. The panels show all spectral elements of the measured covariance matrix $\matV_{QP}$ (blue). The spectra are compared with model predictions assuming parameters from Tab.~\ref{tab:par}, with (solid red) and without (dashed red) contribution from phase noise.  The variance $\sigma_\theta^2$ is obtained by fitting simultaneously the four spectra using Eq.~(\ref{eq:covtheta}) with $\sigma_\theta^2$ as the only free parameter.}
    \label{Fig1}
\end{figure*}

\emph{Analysis} - Having established the system parameters we now focus on the characteristics of the quadratures of the optical field measured at the output of the cavity. We are interested in observing squeezed light generated by the interaction with the levitated nanoparticle, that is, field fluctuations below the shot noise level. Typically, this is verified by experimentally implementing a homodyne detection, which allows to analyze the fluctuations of a chosen field quadrature. In our case, however, the necessary phase sensitive detection is reconstructed in post processing starting from time traces of the BHD. We implement a numerical lock-in amplifier that allows us to measure and stabilize the phase of the beat note in the heterodyne signal.  This approach allows us to reconstruct the complete spectral covariance matrix $\matV_{QP}$ with elements $\mathrm{V}_{ij} = 0.5\langle \{\mathrm{u}_i , \mathrm{u}_j\} \rangle$ where $\mathbf{u} = \big(Q_m(\omega),\,  P_m(\omega)\big)$. Here $Q_m$ and $P_m$ are the measured field quadratures in the frequency domain, normalized to the vacuum level, and $\{\cdot,\cdot \}$ indicates the anti-commutator. A few considerations are in order: (\emph{i}) the determination of $\matV_{QP}$ is completely model independent; (\emph{ii}) the phase references which determine the quadratures, both in the model and in $\matV_{QP}$, are arbitrary and are chosen to match; (\emph{iii}) the simultaneous measurement of the conjugate variables $(Q_m,P_m)$ entails a noise penalty which in this context can be interpreted as an additional factor of 1/2 in the total detection efficiency $\eta$~\cite{Bowen2015}.

\emph{Results} - We show in Fig.~\ref{Fig1} the measured power spectral densities (PSD) composing the elements of the matrix $\matV_{QP}$. The total measurement time was $250$\,s, with a detection efficiency of $\eta=0.32$ (without the additional factor $1/2$ mentioned above). Interestingly, the shape of the diagonal elements, i.e., the PSD of the quadratures $S_{Q_mQ_m}$ and $S_{P_mP_m}$, differ significantly from a pair of Lorentzian peaks. This is a direct consequence of the strong coupling regime, where the mechanical modes are hybridized with the optical one. Indeed, it can be shown that, when the cavity detuning is scanned across the mechanical modes resonances,  the parameters of Tab.~\ref{tab:par} lead to a double avoided crossing~\cite{Ranfagni2021Vectorial,dare2024ultrastrong}.

Comparison between the experiment and the calculated PSDs (see \cite{Deplano2025highpurity} and Appendix~\ref{appendix}) using the parameters obtained in Ref.~\cite{Deplano2025highpurity}, however, shows small discrepancies. This is more evident for the off-diagonal PSDs, while good agreement remains along the diagonal. This behavior leads us to conjecture a residual phase noise affecting the determination of the optical quadratures. We model it by considering a fluctuating phase $\theta$, Gaussian-distributed with variance $\sigma_\theta^2$, over which the analytical covariance matrix needs to be averaged. To this end, we consider a rotation of the covariance matrix as $\matV_{QP}^\theta(\theta)=\mathbf{R}\matV_{QP}\mathbf{R}^T$, where $\mathbf{R}$ is a rotation matrix, and average over realizations of $\theta$ according to

\begin{equation*}
    \tilde{\matV}_{QP}=\frac{1}{\sqrt{2\pi}\sigma_\theta}\int_{-\infty}^\infty \matV_{QP}^{\theta}(\theta) \,\text{exp} \bigg(  -\frac{\theta^2}{2 \sigma_\theta^2} \bigg)  d\theta. 
\end{equation*}
\noindent It can be explicitly written as
\begin{equation}\label{eq:covtheta}
\begin{split}
    &\tilde{\matV}_{QP}=e^{-\sigma_\theta^2} \, \times \\  
&\begin{pmatrix}
S_{PP}\sinh(\sigma_\theta^2) + S_{QQ}\cosh(\sigma_\theta^2) &
S_{QP}\cosh(\sigma_\theta^2) - S_{PQ}\sinh(\sigma_\theta^2) \\
S_{PQ}\cosh(\sigma_\theta^2) - S_{QP}\sinh(\sigma_\theta^2) &
S_{QQ}\sinh(\sigma_\theta^2) + S_{PP}\cosh(\sigma_\theta^2)
\end{pmatrix}.
\end{split}
\end{equation}
We fit the experimental data with Eq.~ (\ref{eq:covtheta}) to determine the unknown variance $\sigma_\theta^2$, which is the only free parameter. All other parameters necessary for the evaluation of the model have been fixed to the nominal values in Tab.~\ref{tab:par}.  As shown in Fig.~\ref{Fig1}, the agreement drastically improves, supporting the phase noise assumption. The fit returns a variance value of $\sigma_\theta^2= 0.062\,\mathrm{rad}^2$.
 
\begin{figure*}[h]
    \centering
    \includegraphics[width=\textwidth]{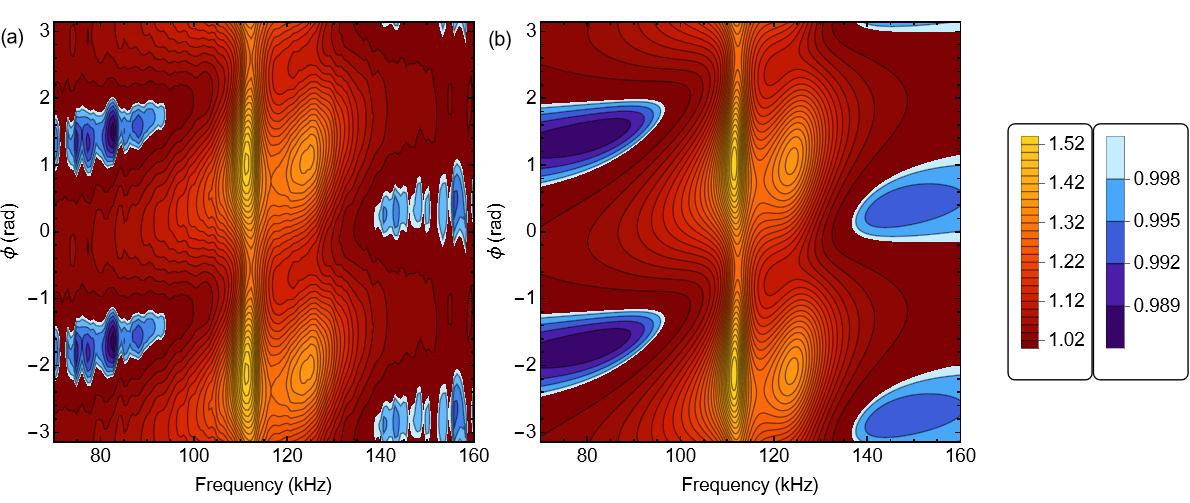}
    \caption{2D map of the optical quadrature spectrum. (a) experimental, model independent map. (b) model prediction of  $\tilde{S}_Q^\phi$ including a phase noise contribution. In both maps, the color scale has been splitted into $\tilde{S}_Q^\phi>1$ (red scale) and $\tilde{S}_Q^\phi<1$ (blue scale) to highlight regions showing sub-shot noise statistics. Two of such regions are evident, below the mechanical resonances (around $80$\,kHz) and above them (around $150$\,kHz). The corresponding detection phases are separated by $\simeq\pi/2$.}
    \label{Fig2}
\end{figure*}

Having reconstructed the covariance matrix $\tilde{\matV}_{QP}$ it is straightforward to obtain the quadrature spectrum $\Tilde{}{S}_Q^\phi$ for an arbitrary detection phase $\phi$. This can be written in simple form, even in the presence of phase noise, as

\begin{equation}\label{eq:sth}
        \tilde{S}_Q^\phi=
        \frac{S_{QQ}+S_{PP}}{2}+\frac{S_{QQ}-S_{PP}}{2\,}\,e^{-2\sigma_\theta^2}\,
\cos(2\phi)+S_{QP}\,e^{-2\sigma_\theta^2}\,\sin (2\phi).
\end{equation}
Using Eq.~(\ref{eq:sth}) we can construct a 2D map showing the quadrature spectrum in a frequency bandwidth accross the \textit{x} and \textit{y} mode resonances, as a function of the phase $\phi$, akin to a tomographic homodyne measurement. The resulting experimental map is shown in Fig.~\ref{Fig2}(a) where it is also compared with the model prediction (Fig.~\ref{Fig2}(b)). The color code in the maps is chosen to clearly separate regions with sub-shot noise statistics (blue scale). Two such regions are evident, in agreement with the model predictions.  
 
\begin{figure}[h]
    \centering
    \includegraphics[width=8.6cm]{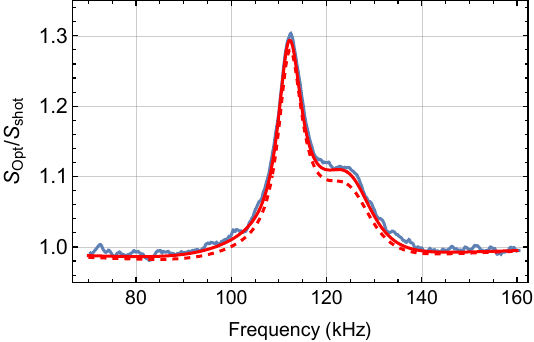}
    \caption{Optimal quadrature spectrum. The optimal experimental spectrum (blue) is compared with model predictions both with (solid red) and without (dashed red) phase noise contribution. Squeezed fluctuations are observable both below and above the mechanical resonances, reflecting the two regions evident in Fig.~\ref{Fig2}. While the inclusion of phase noise in the model drastically improve the agreement with the experimental spectrum, its effect is almost negligible in the squeezed regions.  }
    \label{Fig3}
\end{figure}

A more direct and quantitative way to condense the information displayed in Fig.~\ref{Fig2} is to evaluate the optimal squeezing spectrum $\tilde{S}_Q^{\text{opt}}$, where the detection phase $\phi$ is optimized at every frequency~\cite{pontin2014Frequencynoise}. It can be written as

\begin{equation}
    \begin{split}
        \tilde{S}_Q^{\text{opt}}(\omega)=&S_{QQ}(\omega)+S_{PP}(\omega)-e^{-2\sigma_\phi^2}\\
        &\times \sqrt{[S_{QQ}(\omega)-S_{PP}(\omega)]^2+4[S_{QP}(\omega)]^2}.
    \end{split}
\end{equation}
The resulting optimal spectrum is shown in Fig.~\ref{Fig3}, where it is compared with model predictions. As for the quadratures, without considering the effect of phase noise (dashed line in Fig.~\ref{Fig3}) some discrepancies with the experimental spectrum emerge. These are particularly evident around the peak at $\simeq125$\,kHz. The agreement markedly improves when phase noise is included. However, in the regions where squeezing is observed, i.e., around $80$\,kHz and $150$\,kHz, phase noise has a negligible effect. We point out that such remarkable agreement for the optimal spectrum implies that the agreement is quantitatively similar for any quadrature, at arbitrary detection phase $\phi$. 

\emph{Discussion} - The maximum observed squeezing is mild, but the spectrum is consistently below the shot noise between $70$\,kHz and $95$\,kHz, with a lowest variance for a measured quadrature of $\simeq0.98$, i.e., $2\%$ below vacuum fluctuations. However, this is achieved through two oscillators with occupation numbers well below unity. Observation of squeezing requires the optical field backaction to provide a significant contribution to the total motional variance in order for the correlations between quadratures to emerge. This can be quite challenging to achieve without significantly heating the motion. Furthermore, one of the main limits remains the overall detection efficiency. Tracing back to the cavity output the ideally available squeezing would double to $4\%$, which means that homodyne detection in combination with a variational approach~\cite{Kampel2017Improving} can prove advantageous, at the cost of providing a less comprehensive measurement. An additional factor 2 improvement can be obtained changing the cavity configuration from symmetric to single ended, thus reaching $\sim8\%$.  Finally, in the experimental results presented here collisions with background gas still represent a relevant fraction of the total heating rate. Thus the observable squeezing can potentially be further increased by reaching the photon recoil limit. To move past this barrier, a more sophisticated approach will be necessary.  For example, adding a weak probe field would allow to implement a quantum non-demolition measurement of the field fluctuations~\cite{pontin2018Quantum}. 

In conclusion, we have demonstrated for the first time optical squeezing mediated by two mechanical oscillators simultaneously cooled to their quantum ground state, with occupation numbers well below unity. Our results represent a significant advance in cavity optomechanics, bridging the gap between purely mechanical quantum control and non-classical optical fields.  Even if the measured squeezing is not very pronounced, several pathways for improvement are evident. Enhanced detection efficiency through improved optical collection and single-ended cavity configurations could increase the observed squeezing by a factor of 4 or more. Reducing mechanical heating, by improving vacuum conditions, would allow access to even lower occupation numbers and a larger degree of squeezing. The implementation of quantum non-demolition measurements and variational readout schemes could further enhance the observable squeezing while reducing noise penalties inherent in heterodyne detection.

Looking forward, our platform provides a natural testbed for exploring entanglement between multiple mechanical modes~\cite{bose1997Preparation,vitali2007Optomechanical,paternostro2007Creating}, optomechanical quantum state transfer~\cite{Palomaki2013Coherent,Weaver2017Coherent}, and novel approaches to quantum metrology~\cite{Li2025QuantumEnhanced}. The combination of strong coupling, quantum ground state occupation of multiple modes, and excellent control over the mechanical and optical parameters positions levitated optomechanics as a powerful system for exploring quantum mechanics at the mesoscopic scales.

We acknowledge financial support from PNRR MUR Project No. PE0000023-NQSTI and by the
European Commission-EU under the Infrastructure I-PHOQS “Integrated Infrastructure Initiative
in Photonic and Quantum Sciences ” [IR0000016, ID D2B8D520, CUP D2B8D520].

\bibliographystyle{apsrev4-2} 
\bibliography{main_bib}

\appendix
\begin{center}
    \section{Theoretical model}\label{appendix}
\end{center}

The optomechanical model describing the motion of the nanosphere in the transverse tweezer plane, coupled to a cavity mode, is described in Ref. \cite{Deplano2025highpurity}. In particular, the Langevin equations for the amplitude quadrature of the intracavity field and the two mechanical modes, written in the Fourier space, are given in Eqs. (7-9) of its Supplementary Information, and the equation for the phase quadrature $\Pc$ can be similarly obtained from Eq. (6). We report them below:   
\begin{eqnarray}
\Qc(\omega) &=& i \left(\chic(\omega) - \chic^*(-\omega) \right) \left(g_x\, x(\omega)+g_y\, y(\omega)   \right)+\sqrt{\kappa}\left( \chic(\omega)\, a_{\mathrm{in}} + \chic^*(-\omega)\, a^{\dagger}_{\mathrm{in}} \right)
\label{eq_Lang1} \\
\Pc(\omega) &=& \left(\chic(\omega) + \chic^*(-\omega) \right) \left(g_x\, x(\omega)+g_y\, y(\omega)   \right)+i \sqrt{\kappa}\left( -\chic(\omega)\, a_{\mathrm{in}} + \chic^*(-\omega)\, a^{\dagger}_{\mathrm{in}} \right)
\label{eq_Lang2} \\
x(\omega) &=&  2  \chi_x(\omega) \left(g_x\, Q(\omega) + \sqrt{\Gamma_x}\, \xi_x \right) 
\label{eq_Lang3} \\
y(\omega) &=&  2  \chi_y(\omega) \left(g_y\, Q(\omega) + \sqrt{\Gamma_y}\, \xi_y \right)  \, .
\label{eq_Lang4} 
\end{eqnarray}
Here $x$ and $y$ are dimensionless position operators normalized to their respective zero-point fluctuations $\sqrt{\hbar/2 m \Omega_j}$ ($m$ is the mass of the nanosphere, $\Omega_j$ with $j= (x, y)$ the oscillation frequencies in the absence of optomechanical interaction), $g_j$ are the optomechanical coupling rates, $\Gamma_j$ the heating rates, and $\kappa$ is the cavity width.  
Fourier transformed ($\mathcal{F}$) operators are defined as $O\left(\omega\right)\equiv\mathcal{F}\left[O \left(t\right)\right]=\int O(t) e^{i\omega t} \ud t$, and
$O^{\dagger}\left(\omega\right)=\mathcal{F}\left[O^{\dagger}\left(t\right)\right]$, while the complex conjugate of a complex function $f(\omega)$ is denoted as $f^{*}(\omega)$. In the following, we will use power spectral densities (PSD) of the operators, defined as $\,S_{OO}=\int \langle O^{\dagger}(\omega')\,O(\omega)\rangle\frac{\ud \omega'}{2\pi}$, cross spectral densities defined as $\,S_{O_1 O_2}=\int \langle O_1^{\dagger}(\omega')\,O_2(\omega)\rangle\frac{\ud \omega'}{2\pi}$, and symmetrized spectra $\,\bar{S}_{O_1 O_2} = 0.5 \left(S_{O_1 O_2}(\omega)+S_{O_1 O_2}(-\omega)\right)$. 
We have also defined the mechanical and optical susceptibilities
\begin{eqnarray}
\chi_j\left(\omega\right) &=& \frac{\Omega_j}{\Omega_j^2-\omega^2}\\
\chic\left(\omega\right) &=& \frac{1}{-i
\left(\Delta+\omega\right)+\kappa/2}
\end{eqnarray}
where $\Delta$ is the detuning between the tweezer radiation frequency and the cavity mode. 
The input noise sources have spectral densities $\,S_{\xi_i \xi_i} = S_{a_{\mathrm{in}}^{\dagger} a_{\mathrm{in}}^{\dagger}} = 1$. 

By replacing Eqs. (\ref{eq_Lang3}, \ref{eq_Lang4}) into Eqs. (\ref{eq_Lang1}, \ref{eq_Lang2}) and using the input-output relations $Q = \sqrt{\kappa}\,\Qc-\left( a_{\mathrm{in}} + a_{\mathrm{in}}^{\dagger} \right)$ and $P = \sqrt{\kappa}\,\Pc - i \left( - a_{\mathrm{in}} + a_{\mathrm{in}}^{\dagger} \right)$, we can write the output field quadratures as 
\begin{eqnarray}
Q(\omega) &=& \mathcal{A}_{\scriptscriptstyle{Q}}(\omega)\,N + \mathcal{B}_{\scriptscriptstyle{Q}} (\omega)\, a_{\mathrm{in}} +  \mathcal{B}^{*}_{\scriptscriptstyle{Q}} (-\omega)\, a^{\dagger}_{\mathrm{in}} \\
P(\omega) &=& \mathcal{A}_{\scriptscriptstyle{P}}(\omega)\,N + \mathcal{B}_{\scriptscriptstyle{P}} (\omega)\, a_{\mathrm{in}} +  \mathcal{B}^{*}_{\scriptscriptstyle{P}} (-\omega)\, a^{\dagger}_{\mathrm{in}} 
\end{eqnarray}
where we have defined
\begin{eqnarray}
    \mathcal{A}_{\scriptscriptstyle{Q}}(\omega) &=& 
    \frac{i\sqrt{\kappa}\,\left(\chic(\omega) - \chic^*(-\omega)\right)}{1-2 i \left(\chic(\omega) - \chic^*(-\omega)\right) \left( g_x^2 \,\chi_x + g_y^2\,\chi_y \right)}  \\
     \mathcal{B}_{\scriptscriptstyle{Q}}(\omega) &=& 
     \frac{\kappa\,\chic(\omega)}{1-2 i \left(\chic(\omega) - \chic^*(-\omega)\right) \left( g_x^2 \,\chi_x + g_y^2\,\chi_y \right)}-1 \\
     \mathcal{A}_{\scriptscriptstyle{P}}(\omega) &=& 
    \frac{i\sqrt{\kappa}\,\left(\chic(\omega) + \chic^*(-\omega)\right)}{1-2 i \left(\chic(\omega) - \chic^*(-\omega)\right) \left( g_x^2 \,\chi_x + g_y^2\,\chi_y \right)}  \\
     \mathcal{B}_{\scriptscriptstyle{P}}(\omega) &=& 
     -i \left(\frac{\kappa\,\chic(\omega)\,\left(1+4\,i\,\chic^*(-\omega)\left( g_x^2 \,\chi_x + g_y^2\,\chi_y \right) \right)}{1-2 i \left(\chic(\omega) - \chic^*(-\omega)\right) \left( g_x^2 \,\chi_x + g_y^2\,\chi_y \right)}-1 \right)\\
     N &=& 2 \left(g_x\, \chi_x\, \sqrt{\Gamma_x}\,\xi_x+g_y\, \chi_y\, \sqrt{\Gamma_y}\,\xi_y\right)  \, .
\end{eqnarray}
The symmetrized spectra of the output field quadratures read
\begin{eqnarray}
\nonumber
    \bar{S}_{\scriptscriptstyle{QQ}}(\omega)  &=& 4\left(g_x^2\,|\chi_x|^2\,\Gamma_x +  g_y^2\,|\chi_y|^2\,\Gamma_y \right)\,|\mathcal{A}_{\scriptscriptstyle{Q}}(\omega)|^2 \\  &&+ \,0.5\, \left( |\mathcal{B}_{\scriptscriptstyle{Q}}(-\omega)|^2 + |\mathcal{B}_{\scriptscriptstyle{Q}}(\omega)|^2 \right) \\
    \nonumber
     \bar{S}_{\scriptscriptstyle{PP}}(\omega)  &=& 4  \left(g_x^2\,|\chi_x|^2\,\Gamma_x +  g_y^2\,|\chi_y|^2\,\Gamma_y \right)\,|\mathcal{A}_{\scriptscriptstyle{P}}(\omega)|^2 \\ &&+ \,0.5\, \left( |\mathcal{B}_{\scriptscriptstyle{P}}(-\omega)|^2 + |\mathcal{B}_{\scriptscriptstyle{P}}(\omega)|^2 \right)  \\
         \nonumber
     0.5 \left(\bar{S}_{\scriptscriptstyle{QP}}(\omega)+\bar{S}_{\scriptscriptstyle{PQ}}(\omega) \right) &=& 4\left(g_x^2\,|\chi_x|^2\,\Gamma_x +  g_y^2\,|\chi_y|^2\,\Gamma_y \right)\,\mathrm{Re}\left[\mathcal{A}^{*}_{\scriptscriptstyle{Q}}(\omega) \mathcal{A}_{\scriptscriptstyle{P}}(\omega) \right]  \\
     &&+ \,0.5\, \mathrm{Re}\left[\mathcal{B}^{*}_{\scriptscriptstyle{Q}}(-\omega) \mathcal{B}_{\scriptscriptstyle{P}}(-\omega) + \mathcal{B}^{*}_{\scriptscriptstyle{Q}}(\omega) \mathcal{B}_{\scriptscriptstyle{P}}(\omega) \right]  \, .
\end{eqnarray}

The quantum efficiency $\eta$ is taken into account by modifying the spectra as follows: $\,\bar{S}_{\scriptscriptstyle{QQ}} \to \eta \bar{S}_{\scriptscriptstyle{QQ}} + (1-\eta)$, $\,\bar{S}_{\scriptscriptstyle{PP}} \to \eta \bar{S}_{\scriptscriptstyle{PP}} + (1-\eta)$, $\,\bar{S}_{\scriptscriptstyle{QP}} \to \eta \bar{S}_{\scriptscriptstyle{QP}}$, $\,\bar{S}_{\scriptscriptstyle{PQ}} \to \eta \bar{S}_{\scriptscriptstyle{PQ}}$.

In the main text we have used the simplified notation $0.5 \left( \bar{S}_{\scriptscriptstyle{QP}}+\bar{S}_{\scriptscriptstyle{PQ}} \right) \to S_{QP}$, $\bar{S}_{\scriptscriptstyle{QQ}} \to S_{QQ}$ and $\bar{S}_{\scriptscriptstyle{PP}} \to S_{PP}$.

\end{document}